\documentclass[pre,aps,reprint,twocolumn,footinbib,superscriptaddress,longbibliography,floatfix]{revtex4-1}
\usepackage{graphicx}% Include figure files
\usepackage{dcolumn}% Align table columns on decimal point
\usepackage{bm}% bold math
\usepackage{amssymb}
\usepackage{pifont}
\usepackage{amsmath}
\usepackage{times}
\usepackage[nooneline]{subfigure}

\usepackage[spanish]{todonotes}

\begin{document}

\title{Synchronization scenarios in the Winfree model of coupled oscillators}
\author{Rafael Gallego}
\affiliation{Departamento de Matem\'aticas, Universidad de Oviedo, Campus de
Viesques, 33203 Gij\'on, Spain}
\author{Ernest Montbri\'o}
\affiliation{Center for Brain and Cognition. Department of Information and Communication Technologies,
Universitat Pompeu Fabra, 08018 Barcelona, Spain}
\author{Diego Paz\'o}
\affiliation{Instituto de F\'{i}sica de Cantabria (IFCA), CSIC-Universidad de
Cantabria, 39005 Santander, Spain }

\date{\today}

\begin{abstract}
% Macroscopic synchronization is a collective state that usually emerges in
% populations of oscillators. 
The emergence of collective synchronization was reproduced long ago by Winfree
in a classical model consisting of an ensemble of pulse-coupled phase oscillators.
By means of the Ott-Antonsen ansatz, we derive an exact low-dimensional
representation which is exhaustively investigated 
for a variety of pulse types and phase response
curves (PRCs).  Two structurally different
synchronization scenarios are found, which are linked 
via the mutation of a Bogdanov-Takens point. From our results, we infer a general rule
of thumb relating pulse shape and PRC offset with each scenario.  Finally, we
compare the exact synchronization threshold with the prediction of the
averaging approximation given by the Kuramoto-Sakaguchi model.
At the leading order, the discrepancy appears to behave as an odd function of the PRC offset.
\end{abstract}

\maketitle 

%==========================
\section{Introduction}
%==========================
Macroscopic synchronization is a well-known emergent phenomenon arising in 
ensembles of oscillators when, despite their unavoidable differences, 
some fraction of the oscillators spontaneously lock to one another 
and oscillate together with exactly the same frequency~\cite{Win02,Str00,PRK01}.  
Examples of collective synchronization are abundant and surprisingly diverse, 
see e.g.~\cite{Str03}. They include the synchronous flashing of fireflies~\cite{BB76},   
circadian~\cite{liu97} and cardiac~\cite{GM88} rhythms, the spontaneous 
transitions to synchronous stepping~\cite{millenium} and to synchronous 
clapping~\cite{neda00}, or the collective synchronization of 
chemical oscillators~\cite{kiss02}, and arrays of optomechanical cells~\cite{HLQ+11},
and Josephson junctions~\cite{WCS96}.

The first successful attempt to model macroscopic synchronization is due to
Arthur Winfree. In 1967, Winfree proposed a mathematical model consisting of a 
large population of globally coupled oscillators. Assuming weak coupling, 
Winfree postulated the dynamics of the individual oscillators to be well described by a 
single phase variable. Interactions are modeled by means of pulses that 
are emitted by each oscillator and perturb the phase velocity of all the other oscillators.
Mathematically, 
this is expressed
through two independent functions: 
The (infinitesimal) Phase Response Curve (PRC),
determining how the phase of an oscillator changes under perturbations;
and a function specifying for the shape of the pulses. 
Numerical simulations in ~\cite{Win67,Win80} showed that, under suitable conditions, 
the Winfree model displayed a transition from a totally disordered state to collective synchronization,
analogously to phase transition in statistical mechanics.  
Though the Winfree model was later investigated in a few more papers~\cite{AS01,quinn,basnarkov},
the interest soon turned to the simpler and renowned Kuramoto model
~\cite{Str00,PRK01,Str03,ABP+05}.

A new boost in the theoretical understanding of phase-oscillator populations models
occurred in 2008, when Ott and Antonsen (OA) discovered an exact dimensionality reduction 
of the Kuramoto model, called OA ansatz~\cite{OA08,OA09,OHA11}. The discovery of the
OA ansatz opened up the possibility of tackling unresolved problems 
and investigate novel variants and extensions of the Kuramoto model,
see e.g.~\cite{antonsen08,PR08,CS08,AMS+08,AO09,PM09,MBS+09,LOA09,LCT10,alonso11,
HS11,MP11,BAO11,OW12,IPM+13,RO14,PDD16,OA17}.
Remarkably, the OA ansatz is also applicable to pulse-coupled oscillators~
\cite{LBS13,SLB14,Lai14,Lai15,MPR15,PM16,okeeffe16,CHC+17} and, in particular, 
to the original Winfree model~\cite{PM14}. This allows to investigate 
synchronization phenomena which are not accessible using Kuramoto-like models. 
Specifically, the advantage of the Winfree model is that permits to investigate 
separately how pulse shape and PRC type influence collective synchronization. 
Note that the PRC of cells, such as neurons \cite{preyer_butera05,tateno07} 
and cardiac cells \cite{kralemann}, can be measured experimentally.
Therefore understanding better the Winfree model should 
contribute to narrow the gap between mathematical models and biological phenomena.

Here we build on our previous work~\cite{PM14}, and systematically analyze the impact 
of (i) pulse shapes and (ii) PRC offsets, onto collective synchronization 
in the Winfree model. We find that the phase diagram obtained in~\cite{PM14} is not unique, 
and that a novel synchronization scenario emerges for certain pulse types
via the mutation of a codimension-two Bogdanov-Takens (BT) point. 
We end investigating the limit in which the oscillators are 
nearly identical and very weakly coupled. In that situation
the averaging approximation is valid, and a Kuramoto-like model captures the dynamics
with a level of accuracy that is measured numerically.

The paper is organized as follows: 
In Sec.~\ref{sec::model} we introduce the Winfree model, and 
in Sec. \ref{sec_oa} an exact reduction to two ordinary differential equations (ODEs)
is derived using the OA ansatz. 
In Sec. \ref{sec::res} we present the results obtained from those two ODEs,
for a variety of pulse shapes and PRCs.
In Sec.~\ref{sec::kuramoto} we compare the Winfree model with
its averaging approximation. 
Finally, in Sec.~\ref{sec::con} we address the conclusions of this work.

%==========================
\section{The Winfree Model}\label{sec::model}
%==========================

The Winfree model~\cite{Win67,Win80} consists of an ensemble of $N\gg1$ 
globally coupled phase oscillators with 
heterogeneous natural frequencies $\omega_i$, $i=1,\ldots,N$.
The phases $\theta_i$ are governed by the set of $N$ ordinary differential equations (ODEs)
\begin{equation}
\dot\theta_i=\omega_i+ Q(\theta_i) \, \frac{\varepsilon}{N} \sum_{j=1}^N
P(\theta_j).
\label{model} 
\end{equation}
Here, each oscillator receives the input of the mean field
$h=N^{-1} \sum_{j=1}^N P(\theta_j),$ 
and its response to it depends on its own phase through the PRC function $Q(\theta)$.
Both $P$ and $Q$ are $2\pi$-periodic functions on the real line, and hence can be 
defined either in the range $[0,2\pi)$ or in the range $[-\pi,\pi)$.
Finally, the global coupling strength is controlled by 
the parameter $\varepsilon>0$.

%%%%%%%%%%%%%%%%%%%%%%%%%%%%%%%%%%%%%%%%%%%%%%%%%%%%%%%%%%%%%%%%%%%%%%
\begin{figure}
\includegraphics[width=65mm,clip=true]{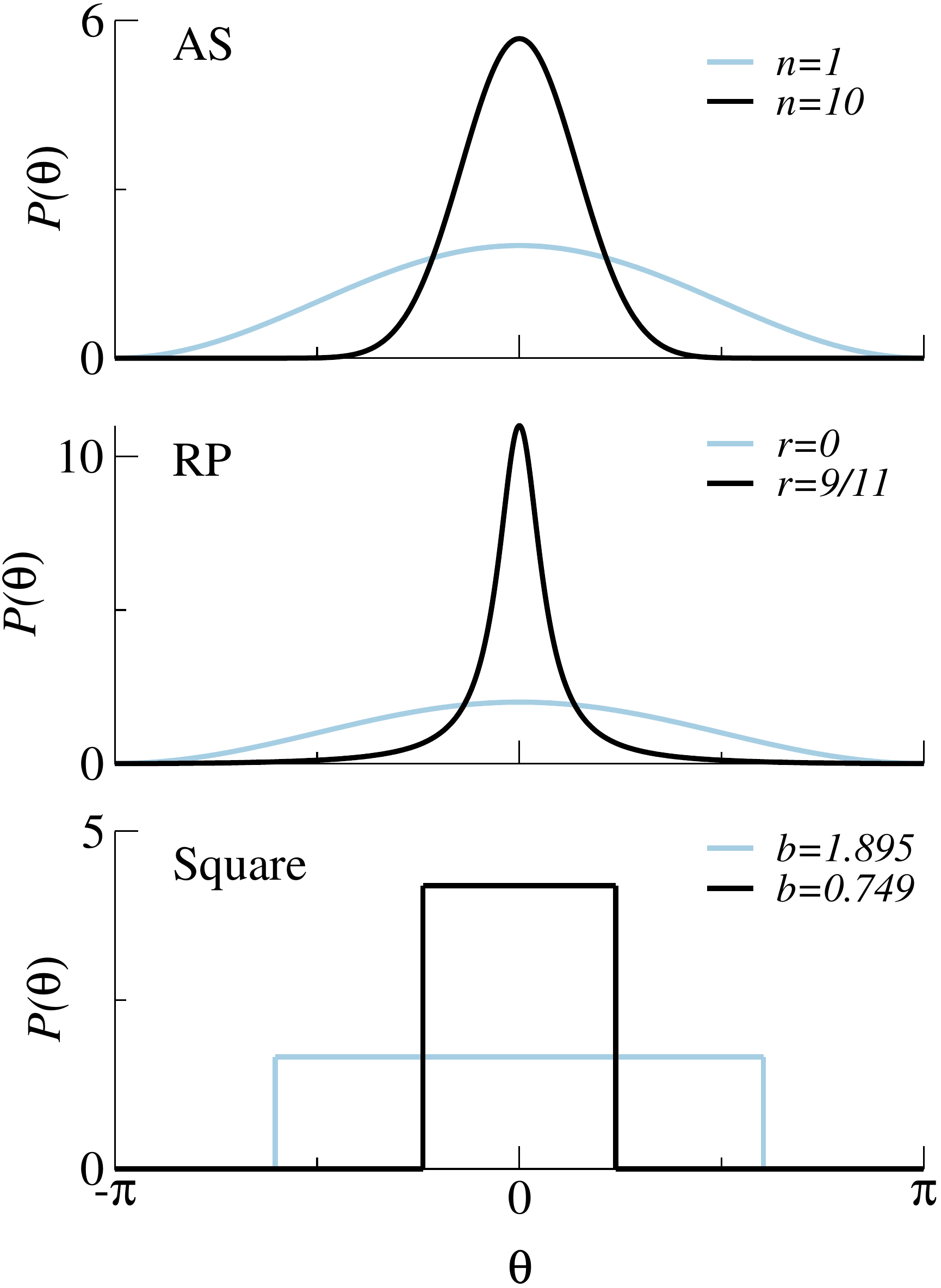}
\caption{The three pulse functions $P(\theta)$
used in this paper. Two different widths are displayed in each panel. 
See Table \ref{table} for the detailed mathematical form of the pulses. 
Functions with the same line style have the same shape factor $\Pi$.}
\label{fig::1}
\end{figure}
%%%%%%%%%%%%%%%%%%%%%%%%%%%%%%%%%%%%%%%%%%%%%%%%%%%%%%%%%%%%%%%%%%%%%%

\subsection{Pulse shape, $P(\theta)$}
\label{sec::pulse}
%%%%%%%%%%%%%%%%%%%%%%%%%%%%%%%%%%%%%

The function $P$ in Eq.~\eqref{model} specifies the form of the pulses. 
We  only consider pulses with the following properties:
\begin{enumerate}
 \item[(i)] $P$ is unimodal and symmetric around $\theta=0$.
 \item[(ii)] $P$ vanishes at $\theta=\pi$.
 \item[(iii)] $P$ has a normalized area: $\int_{-\pi}^\pi P(\theta) d\theta=2\pi$.
 \end{enumerate}
We consider the three pulse types with finite width shown in 
Fig.~1, and defined in Table \ref{table}. 
The first pulse, labeled as AS, was originally adopted by Ariaratnam
and Strogatz \cite{AS01}, and is commonly used in recent studies of pulse 
coupled-phase oscillators~\cite{goel02,PM14,LBS13,SLB14,Lai14,Lai15,okeeffe16,CHC+17}. 
Additionally, we consider a variant of the pulse used by 
O'Keeffe and Strogatz in~\cite{okeeffe16} equal to the Poisson kernel, 
but with an offset so that it fulfills the condition (ii). We term this pulse as ``Rectified Poisson kernel'' (RP).
Finally we consider a square pulse with a flat profile and vanishing in a finite interval of
theta: $[-\pi,-b) \cup(b,\pi)$.

Concerning the macroscopic dynamics of the Winfree model, 
the precise value of $N$ becomes irrelevant provided it is
large enough (i.e. only trivial finite-size fluctuations are observed). However, 
for Dirac delta pulses, this is not the case, as we discuss in Sec.~\ref{sec::dirac}. 
The Dirac delta is the limiting case of the pulse types considered, 
i.e.~$n\to\infty$, $r\to1$, and $b\to0$ for the AS, RP, and square pulses, respectively.

%%%%%%%%%%%%%%%%%%%%%%%%%%%%%%%%%%%%%%%%%%%%%%%%%%%%%%%%%%%%
\begin{figure}
  \includegraphics[width=65mm,clip=true]{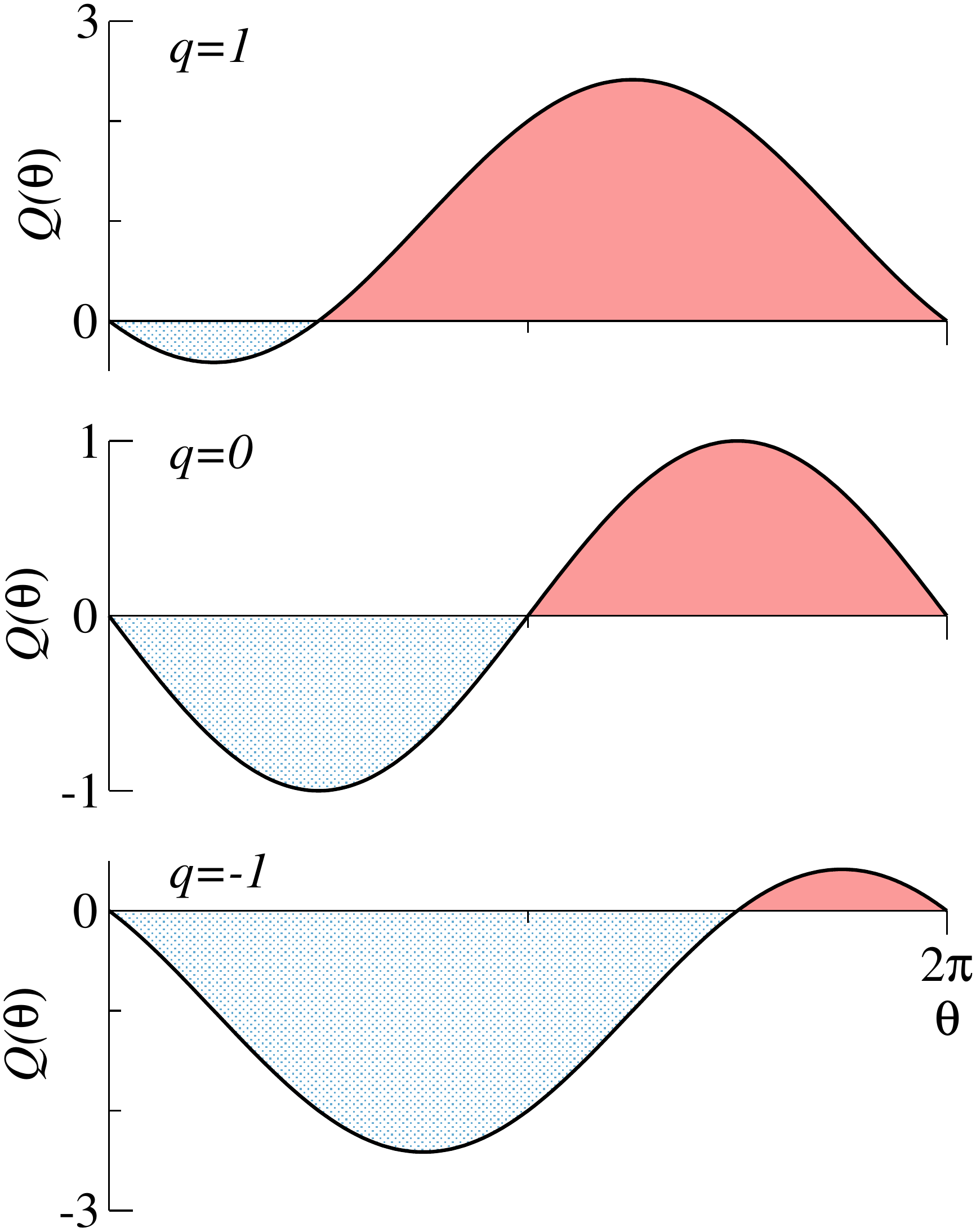}
  \caption{Phase-Response Curves~\eqref{prc} used in this paper.  
  Brief, perturbations lead to either a
  phase delay (light-shaded blue) or a phase advance (shaded red), depending on the state $\theta$
  of the oscillator. The sign and magnitude of parameter $q$ controls
  the offset of the PRC, and hence determines if pulse interactions are mostly
  promoting ($q>0$) or delaying ($q<0$) phase shifts.} 
  \label{fig::qtheta}
\end{figure}

%%%%%%%%%%%%%%%%%%%%%%%%%%%%%%%%%%%%%%%%%%%%%%%%%%%%%%%%%%%%

%%%%%%%%%%%%%%%%%%%%%%%%%%%%%%%%%%%%%%%%%%%%%%%%%%%%%%%%%%%%%%%%%%%%%%%%%%%%%%%%%%%%%%%%%%%%
\begin{table*}
\centering % used for centering table
\begin{tabular}{c c  c  c  c } % centered columns (5 columns)
\hline\hline %inserts double horizontal lines
Pulse name & $P(\theta)$ &  Parameter  & Mean field: $h(Z)$ & Shape factor: $\Pi$ \\ %[0.5ex] % inserts table
%heading
\hline \\ [-2ex]% inserts single horizontal line
Ariaratnam-Strogatz (AS) &  $a_n (1+\cos\theta)^n$ & $n\in\mathbb{Z}^+$ &  $1+
(n!)^2\displaystyle\sum_{k=1}^n \dfrac{Z^k + (Z^*)^k}{(n+k)! (n-k)!}$ &  $\dfrac{n}{n+1}$  \\  \hline 
Rectified-Poisson (RP) & $\dfrac{(1-r)(1+\cos\theta)}{1-2r\cos\theta +r^2}$ & $r\in(-1,1)$ 
 & $\mbox{Re} \left[ \dfrac{1+Z}{1-rZ} \right]$ & $\dfrac{1+r}{2}$ \\ \hline
Square &
$
\begin{cases}
\pi/b \, \mbox{for} \, |\theta|\le b \\
0 \, \mbox{otherwise}
\end{cases}
$
& $b\in(0,\pi)$  & $1-\dfrac{1}{b}\mbox{Im} 
\left[ \ln(1-Z e^{ib})-\ln(1-Z e^{-ib})] \right] $ & $\dfrac{\sin b}{b}$ \\ \hline
Dirac delta & $2\pi\delta(\theta)$ & $-$   & $\mbox{Re} \left[ \dfrac{1+Z}{1-Z} \right]$ & $1$ \\[1ex] % [1ex] adds vertical space
\hline\hline 
\end{tabular}
\caption{Summary of the various pulse functions considered. 
The normalizing constant of the AS pulse is $a_n=2^n (n!)^2 /(2n)!=n!/(2n-1)!!$.
The fourth column shows the mean field $h(Z)$, 
which is the function entering in Eq.~\eqref{Z} describing 
the system's 
mean field 
dynamics exactly. 
In the last column, the shape factor $\Pi$ 
quantifies the effective strength
of each pulse under the averaging approximation, see Eq.~\eqref{av_winfree}.
In Fig.~\ref{fig::1}, lines of the same style share the same $\Pi$ value.}
\label{table} 
% \comentarioi{Cuidado con lo de \emph{same color}}
\end{table*}
%%%%%%%%%%%%%%%%%%%%%%%%%%%%%%%%%%%%%%%%%%%%%%%%%%%%%%%%%%%%%%%%%%%%%%%%%%%%%%%%%%%%%%%%%

%%%%%%%%%%%%%%%%%%%%%%%%%%%%%%%%%%%%%%%%%%%%%%%%%%%%
\subsection{Phase-response curve (PRC), $Q(\theta)$}
%%%%%%%%%%%%%%%%%%%%%%%%%%%%%%%%%%%%%%%%%%%%%%%%%%%%
The influence of a certain (small) perturbation on the phase of an
oscillator is determined by the PRC, $Q(\theta)$. 
Here we assume that (i) the PRC vanishes at the phase where 
the pulses peak, i.e. at~$Q(\theta=0)=0$; and (ii) the PRC
has a sinusoidal shape. 
This latter condition is crucial, for the OA theory to be applicable.
The constraints (i) and (ii) lead us to the following one-parameter family of PRCs
\begin{equation}
Q(\theta)
=\frac{\sin\beta-\sin(\theta+\beta)}{\cos\beta}
=q(1-\cos\theta)-\sin\theta,
\label{prc}
\end{equation}
where parameter $q=\tan\beta$ determines the degree of asymmetry of the PRC.
As illustrated in Figure~\ref{fig::qtheta} , 
$Q$ is more positive (advancing) than negative for $q>0$, while it is
more  negative (retarding) for $q<0$. The case $q=0$ corresponds to 
a perfectly balanced PRC. Hence, we call $q$ `offset parameter' hereafter.
Note that  in \cite{PM14} the PRC is defined in a slightly different manner:
Here $\varepsilon$ is equivalent to $\varepsilon \cos\beta$ in 
our previous work.

%%%%%%%%%%%%%%%%%%%%%%%%%%%%%%%%%%%%%%%%%%%%%%%%
\subsection{Frequency distribution, $g(\omega)$}
%%%%%%%%%%%%%%%%%%%%%%%%%%%%%%%%%%%%%%%%%%%%%%%%

As indicated above, heterogeneity in the population enters through the set of natural
 frequencies $\omega_i$. 
As we show in the next section, to simplify the analysis of the Winfree model~\eqref{model}
it is convenient to adopt a Lorentzian distribution centered at $\omega_0$ 
with half-width $\Delta$:
\begin{equation}
g(\omega)= \frac{\Delta/\pi}{(\omega-\omega_0)^2+\Delta^2}.
\label{lorentzian}
\end{equation}
%

%==========================
\section{Low-dimensional dynamics of the Winfree model}\label{sec_oa}
%==========================

In the following we present a reduction of the dimensionality of
our problem using the so-called Ott-Antonsen theory, 
what permits to determine the system dynamics 
exactly in the thermodynamic limit $N\to\infty$ for all parameter values.
Hence, we introduce the density $F$ function, such that
$F(\theta|\omega,t)\, d\theta$ represents the fraction of oscillators with
phases between $\theta$ and $\theta+d\theta$, and
natural frequency $\omega$ at a time $t$.  The conservation of the number of oscillators imposes $F$
to obey the continuity equation:
\begin{equation}
  \partial_t F + \partial_\theta\left(F \, \dot \theta\right)=0
\label{cont}
\end{equation}
To solve this equation it is natural to use a Fourier expansion of $F$:
\begin{equation}
F(\theta|\omega,q,t)=\frac{1}{2\pi}
\left\{ 1+ \left[\sum_{m=1}^{\infty} \alpha_m(\omega,t) e^{im\theta}  + \mbox{c.c.}\right] \right\},
\label{oa}
\end{equation}
where c.c.~stands for complex conjugate.
We adopt the OA ansatz assuming that the $m$-th mode is the $m$ power of the first mode:
$\alpha_m(\omega,t)=[\alpha(\omega,t)]^m$. This drastic reduction of dimensionality
was justified in \cite{PM14} following \cite{OA09,OHA11}, see also \cite{pietras16}.
Now, inserting \eqref{oa} into the continuity equation
\eqref{cont} we get an equation for $\alpha(\omega,t)$:
\begin{equation}
\partial_t \alpha  = -i(\omega+ \varepsilon h q) \alpha
+ \frac{\varepsilon h}{2} \left[(1+iq) -(1-iq)\alpha^2 \right].
\label{alpha} 
\end{equation}
In this equation the mean field
\begin{equation}
h(t)=\int_{-\infty}^\infty g(\omega) 
\int_0^{2\pi} F(\theta|\omega,t) \, P(\theta) \, d\theta \, d\omega
\label{h}
\end{equation}
couples every $\alpha(\omega,t)$ with all others $\alpha(\omega',t)$.

We use the Kuramoto order parameter $Z$ to monitor the macroscopic
dynamics of the system.
It quantifies the amplitude of first Fourier mode of the density $F$, and 
reads
\begin{equation}
Z(t)=\int_{-\infty}^\infty g(\omega)  \int_0^{2\pi} F(\theta,\omega,t) e^{i\theta} \, d\theta \, d\omega. 
\end{equation}
%%%Aqui esto no lo entiendo, Diego:
Under the assumption that the system evolves in the OA manifold:
\begin{equation}
Z^*(t)=\int_{-\infty}^\infty g(\omega)         \,  \alpha(\omega,t) \, d\omega .
\end{equation}
(The asterisk denotes complex conjugation.)
For Lorentzian $g(\omega)$
this integral over the real line can be computed 
by performing an analytical continuation of $\alpha(\omega,t)$ from real $\omega$ into
complex $\omega=\omega_r+i\omega_i$, see \cite{OA08} for details. Closing the integral by a half-circle at infinity
in the lower complex $\omega$ half-plane, permits to apply the
residue's theorem, obtaining
\begin{equation}
Z^*(t)=\alpha(\omega_p,t) ,
\label{za}
\end{equation}
where $\omega_p=\omega_0-i\Delta$ is the simple pole of $g(\omega)$ inside the 
integration contour. 

The exact, low-dimensional form of the Winfree model with Lorentzian 
frequency distribution \eqref{lorentzian} and sinusoidal PRC \eqref{prc}, 
is obtained setting $\omega=\omega_p$ in Eq.~\eqref{alpha}. Then, we obtain 
the complex-valued ODE
\begin{equation}
\dot Z  = (-\Delta +i \omega_0 )  \, Z
+ \frac{\varepsilon h}{2} \left[1-Z^2 - iq \, (1-Z)^2 \right] .
\label{Z} 
 \end{equation}
To use this equation, 
one needs to express the mean-field as $h(Z)$.
With that aim, it is convenient to expand $P$ in Fourier series:
\begin{equation}
 P(\theta)=\sum_{m=-\infty}^{\infty} c_m e^{im\theta}
 \label{prc_fourier}
\end{equation}
with $c_m=c_{-m}\in \mathbb{R}$ and $c_0=1$, because of the 
properties (i) and (iii) 
stipulated in Sec.~\ref{sec::pulse}. 
Inserting Eqs.~\eqref{prc_fourier} and \eqref{oa} into Eq.~\eqref{h} we get
\begin{equation}
h=1+\sum_{k=1}^{\infty} c_k \int_{-\infty}^\infty g(\omega)     \left\{  \left[\alpha(\omega,t)^*\right]^k+\left[\alpha(\omega,t)\right]^k \right\} d\omega ,
\end{equation}
which again can be simplified applying the residue's theorem, 
and recalling Eq.~\eqref{za} allows to express the result in terms of $Z$ only
\begin{equation}
h=1+\sum_{k=1}^{\infty} c_k \left[Z^k+\left(Z^*\right)^k\right].
\end{equation}
This relation permits to achieve, after some algebra, 
the desired expressions of $h(Z)$ for 
the set of pulse types. They are listed in the fourth column of Table \ref{table}. 

Note that, compared to the AS pulse used in 
previous studies~\cite{AS01,goel02,PM14,LBS13,SLB14,Lai14,Lai15,okeeffe16,CHC+17}, 
the RP pulse has the advantage that $h(Z)$ remains a simple function of $Z$,
no matter how much the pulse width is decreased, see Table I. 
Moreover, though the mean field function $h(Z)$ for square pulses is more cumbersome 
than that of the RP pulses, it still permits to investigate the Winfree model 
with pulses of arbitrary small width, without the drawback of dealing with the 
long sums of the AS pulse's mean field.

%%%%%%%%%%%%%%%%%%%%%%%%%%%%%%%%%%%%%%%%%%%%%%%%%%%%%%%%
\begin{figure}
  \includegraphics[width=\columnwidth]{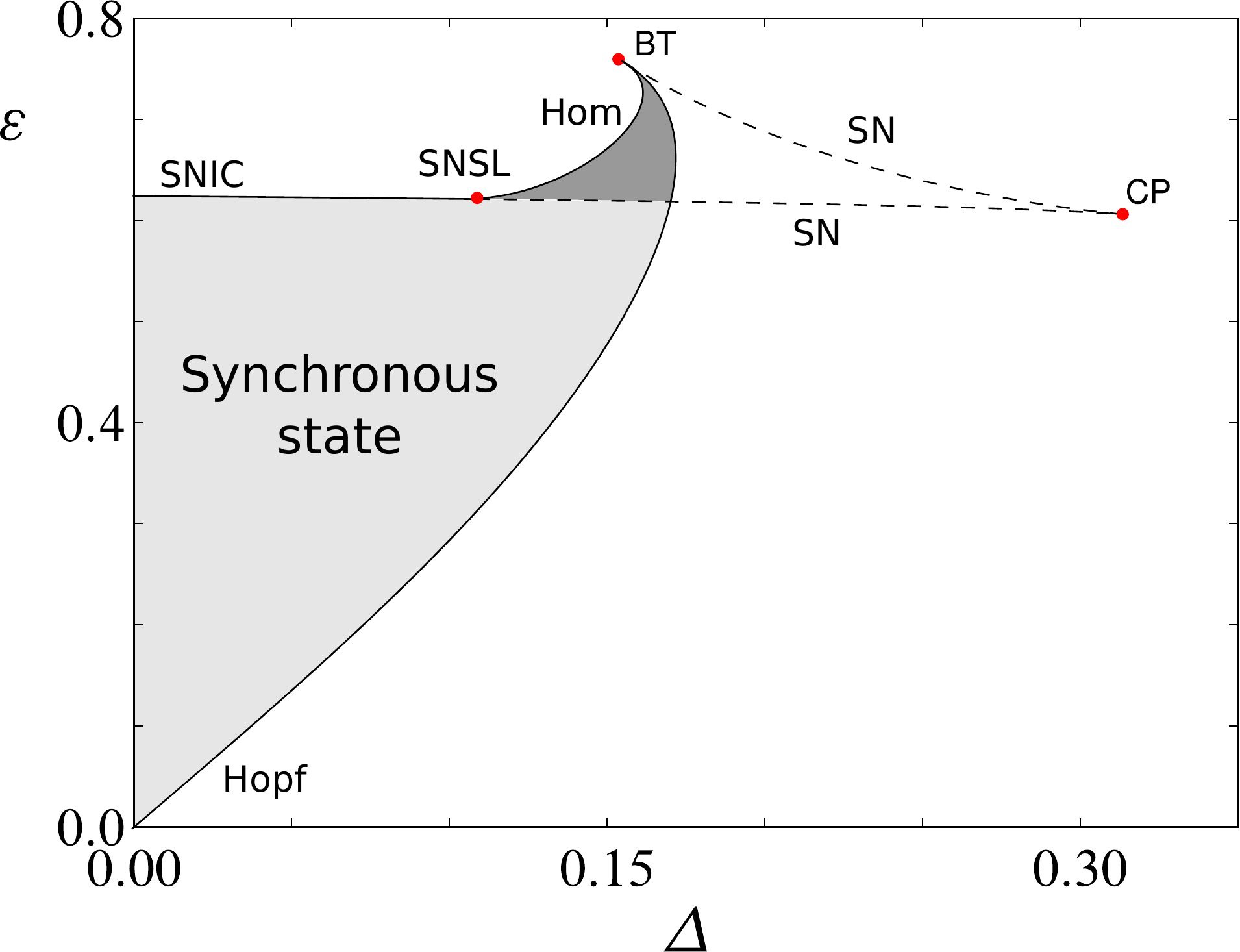}
  \caption{Phase diagram of the Winfree model in the
$(\Delta,\varepsilon)$-plane for a PRC with $q=-1$ and a RP pulse with
$r=0.5$. Bifurcation lines are obtained from Eq.~\eqref{Z}.
In the shaded region there is a stable limit-cycle corresponding
to a synchronized state, see Fig.~\ref{fig::raster}(b).
The boundary of synchronization are Hopf, SNIC and homoclinic bifurcation lines.
In the dark shaded region the limit-cycle (synchronization) coexists with
a stable fixed point (asynchronous state). 
Accordingly the dashed lines are the loci saddle-node bifurcations.
Finally, note that three codimension-two points organize the bifurcation lines:
double-zero eigenvalue Bogdanov-Takens (BT), cusp (CP), and 
saddle-node separatrix-loop (SNSL).}
  \label{fig::pdrp}
\end{figure}
%%%%%%%%%%%%%%%%%%%%%%%%%%%%%%%%%%%%%%%%%%%%%%%%%%%%%%%%%

%==========================
\section{PHASE DIAGRAMS AND PHASE PORTRAITS}
%==========================
\label{sec::res}

Next we exploit the low-dimensional character of Eqs.~\eqref{Z}, to fully investigate the 
bifurcations of the Winfree model for the pulse functions described in 
Table \ref{table}, and various PRC offsets $q$. First note that the dynamics of the 
Winfree  model~\eqref{Z} depends on five parameters:
The coupling strength ($\varepsilon$), the pulse width (through $n$, $r$ or $b$), 
the center and half-width of the frequency distribution ($\omega_0$ and $\Delta$)
and the PRC offset ($q$). From now on, and without lack of generality, 
we set $\omega_0=1$, since this can always be achieved after a trivial rescaling of 
time and parameter $\varepsilon$. 

\subsection{Rectified-Poisson (RP) pulse} \label{sec::rp}
%%%%%%%%%%%%%%%%%%%%%%%%%%%%%%%%%%%%%%%%%%%%%%%%%%%%%%%%%%%%

%%%%%%%%%%%%%%%%%%%%%%%%%%%%%%%%%%%%%%%%%%%%%%%%%%%%%%%%%%%%%
\begin{figure*}
  \includegraphics[width=\textwidth]{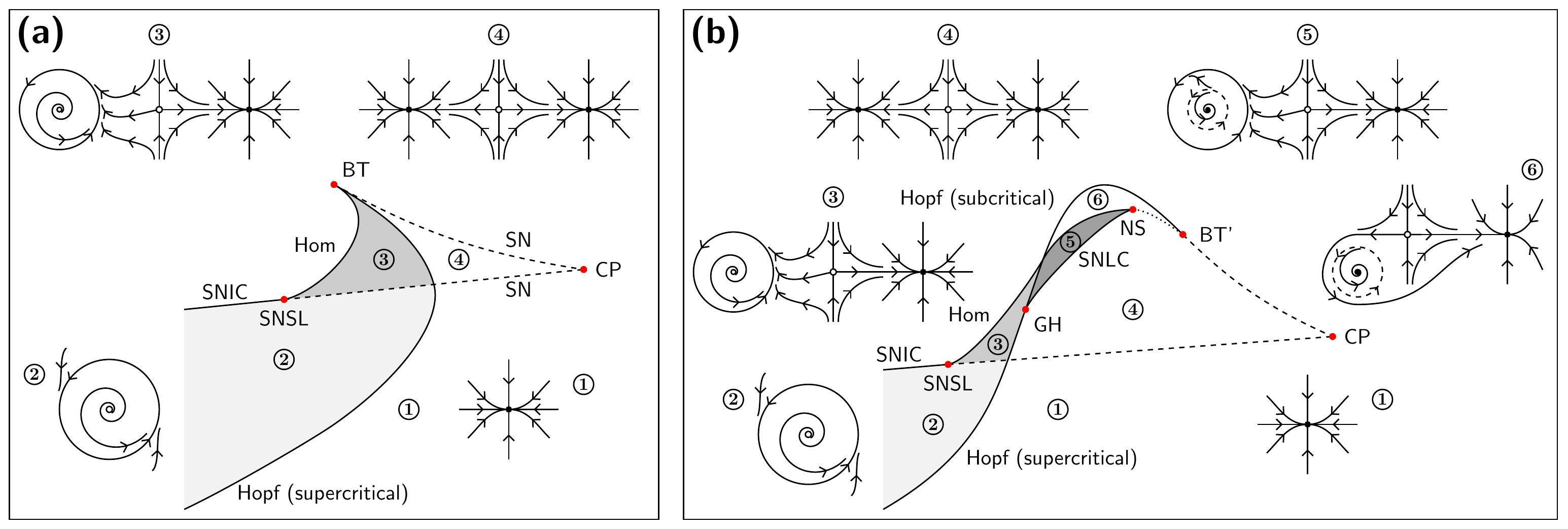}
  \caption{Sketches of the phase diagrams and phase portraits in the different
  regions of the  $(\Delta,\epsilon)$-plane.  Graph (a) displays a typical
  diagram when a BT point comes into play \cite{PM14},
  whereas graph (b) corresponds to the
  case of a BT' point (see text for details). Some details such as the
  transition from stable node to stable focus, or the annihilation of saddle and
  unstable node have been omitted for simplicity.}
  \label{fig::pdsketch}
\end{figure*}
%%%%%%%%%%%%%%%%%%%%%%%%%%%%%%%%%%%%%%%%%%%%%%%%%%%%%%%%%%%%%

Figure \ref{fig::pdrp} shows the phase diagram for the RP pulse ($r=0.5$) with
negative PRC offset ($q=-1$), obtained 
using Eq.~\eqref{Z} with the assistance of the {\sc MatCont} toolbox of {\sc matlab}. 
The diagram is qualitatively identical to those
presented in Ref.~\cite{PM14} for the AS pulse, indicating certain robustness
of the dynamics against modifications of the pulse shape. 
In Fig.~\ref{fig::pdsketch}(a) we
show a sketch of the phase portraits in the regions of interest. 
In the shaded region, labeled $2$, Eq.~\eqref{Z} 
has one attractor of limit-cycle type, meaning that $Z$ exhibits periodic oscillations.
This is reflecting a state of macroscopic synchronization in which a certain part of the population
is self-entrained to a common frequency.  There are three different paths
leading to this state, depending on
which bifurcation line is crossed: Hopf, SNIC
(saddle-node on the invariant circle), or Hom (homoclinic or saddle-loop). Note
that the latter one is a global bifurcation that does not destabilize the
steady state, and in consequence, a region of bistability between synchrony and
asynchrony exists, see the dark shaded region in Figs. \ref{fig::pdrp} and \ref{fig::pdsketch}(a).
Two lines of saddle-node bifurcations of fixed points, emanating from a cusp point (CP),
complete the phase diagram and bound a region of bistability between two stable steady
states (region 4). They  
correspond to two 
asynchronous states 
with a different number of quiescent oscillators. 
For large enough $\Delta$, namely above the CP point, the
fraction of quiescent oscillators varies smoothly
(i.e.~non-hysteretically) with $\varepsilon$.

\begin{figure}
\includegraphics[width=\columnwidth]{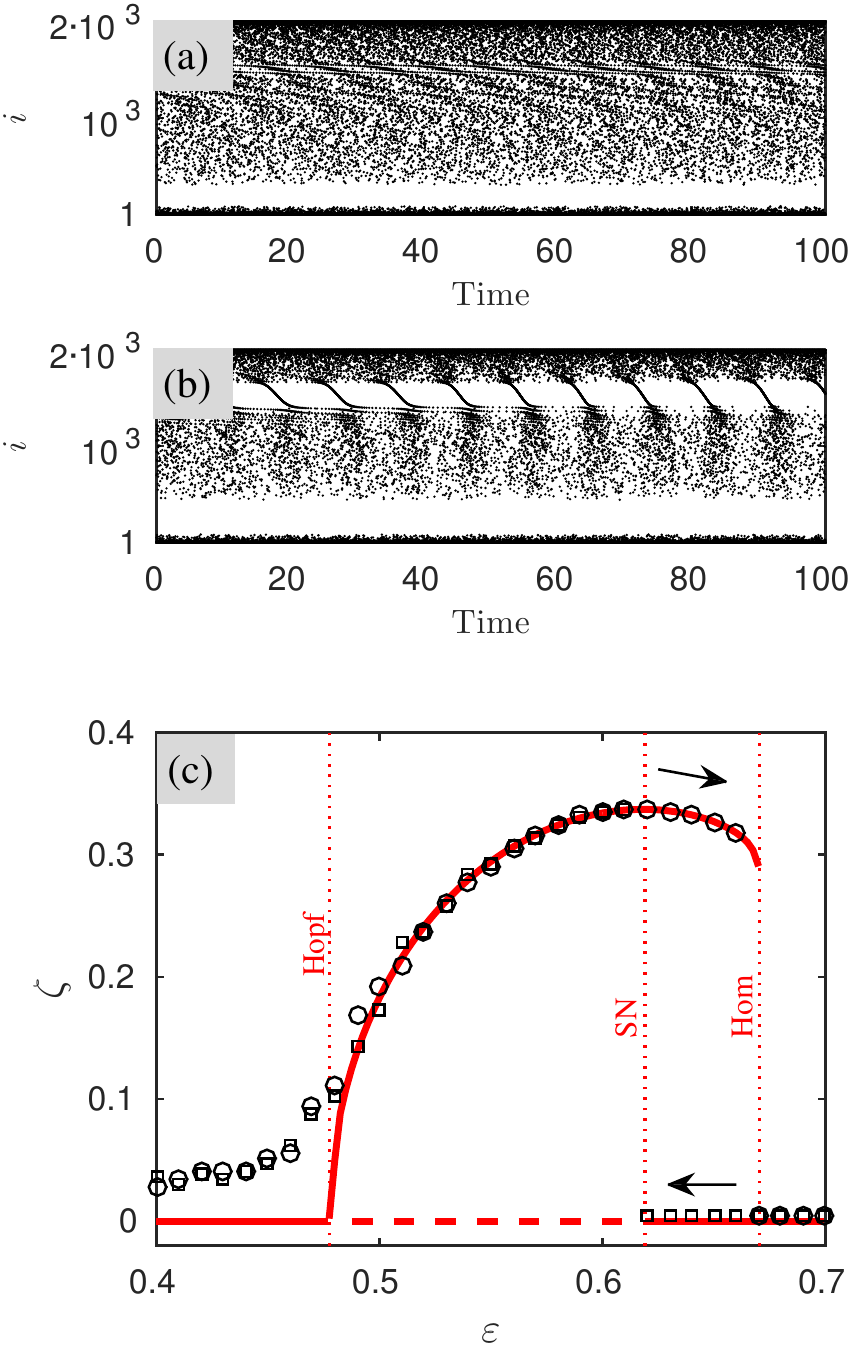}
  \caption{
Panels  (a) and (b) show raster plots of $N=2000$ ``Winfree oscillators'' with $q=-1$ and RP pulses 
with $r=0.5$ ---as in Fig.~\ref{fig::pdrp}. In panels (a) and (b) the coupling strength are $\varepsilon=0.4$,
and $\varepsilon=0.5$, exhibiting incoherent and synchronized states, respectively.
The horizontal white stripe corresponds to oscillators with natural frequencies 
near zero that remain quiescent. The natural frequencies have been deterministically selected 
from a Lorentzian distribution with $\omega_0=1$ and $\Delta=0.15$,
using $\omega_i=\omega_0+\Delta \tan[\pi(2i-N-1)/(2N)]$.
Panel (c): Bifurcation diagram $\zeta$ vs. $\varepsilon$ along the line $\Delta=0.15$. 
The red lines are obtained from the low-dimensional Eq.~\eqref{Z}, and the bifurcations 
(Hopf, saddle-node and homoclinic, from left to right)
are marked by vertical dotted lines. Symbols correspond to numerical simulations of the
Winfree model. Circles (resp.~squares) are the results increasing (resp. lowering) $\varepsilon$.
}
\label{fig::raster}
\end{figure}

To 
verify the 
validity of our analytical results we carried out simulations of the full model
with $N=2000$ oscillators. In Figs.~\ref{fig::raster}(a) and \ref{fig::raster}(b), 
we present raster plots of the incoherent and the synchronized states, respectively 
(a dot is plotted every time an oscillator crosses a multiple of $2\pi$).
Moreover, to be more systematic
we swept parameter $\varepsilon$ along $\Delta=0.15$, i.e. a vertical line in 
Fig.~\ref{fig::pdrp}, with the intention
of testing that the bifurcations were indeed reproduced. As the rotation of the 
oscillators is not uniform, $|Z|$ alone is not a good order parameter to detect bifurcations. 
It is 
more convenient to use the
order parameter proposed by Shinomoto and Kuramoto \cite{ShK86}:
\begin{equation}
\zeta=\overline{|Z-\bar{Z}|} ,
\label{SKOP}
\end{equation}
%%%
where the bar 
means long-time average. For 
asynchronous dynamics, the Shinomoto-Kuramoto order parameter Eq.~\eqref{SKOP} satisfies $\zeta=0$, 
while $\zeta\neq0$ indicates some degree of synchronization.  
Figure~\ref{fig::raster}(c) shows that the 
results of our numerical simulations of the original Winfree model show a good agreement 
with the reduced ODE, Eq.~\eqref{Z}.

%%%%%%%%%%%%%%%%%%%%%%%%%%%%%%%%%%%%%%%%%%%%%%%%%%%%%%%%%%%%%
%%%%%%%%%%%%%%%%%%%%%%%%%%%%%%%%%%%%%%%%%%%%%%%%%%%%%%%%%%%%%

\begin{figure}
  \includegraphics[width=\columnwidth]{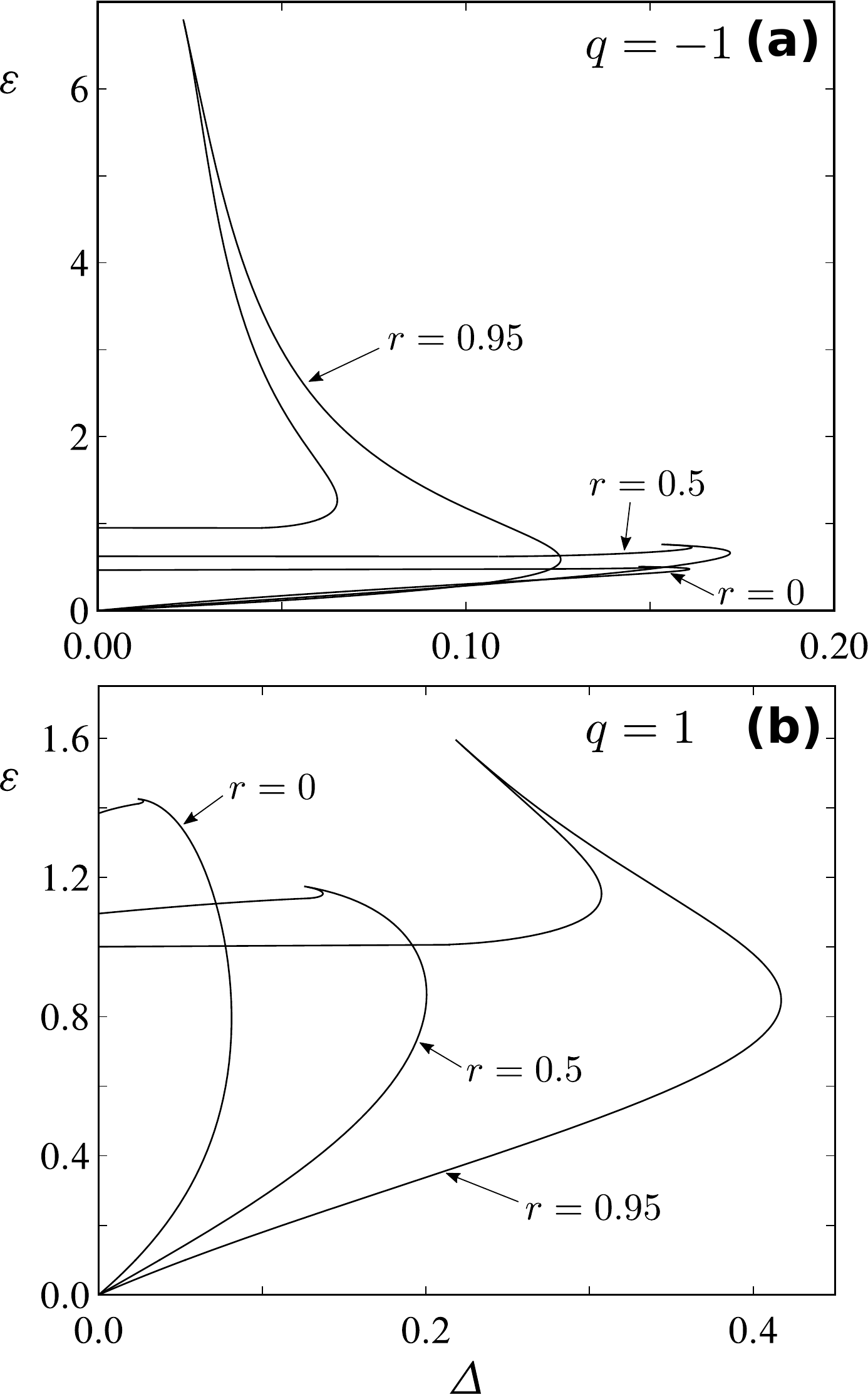}
  \caption{Boundaries of the synchronization region in the
  $(\Delta,\varepsilon)$-plane for the RP pulse and several values of
  $r$.
%% recall that the pulse becomes narrower as $r$ approaches 1. 
  Graphs (a) and (b) correspond to $q=-1$ and $q=1$, respectively.}
  \label{fig::q_rp}
\end{figure}

We next investigate how the synchronization region changes as the pulse width varies.
Figures \ref{fig::q_rp}(a) and \ref{fig::q_rp}(b) show the synchronization boundaries
for $r=0$, $0.5$ and $0.95$, and opposite values of the PRC's offset $q$ ($=\mp1$). 
For the sake of clarity other bifurcation lines have been omitted. 
For positive offsets, 
we find the same result that we found in~\cite{PM14} for AS pulses with 
highly unbalanced PRCs ($q\gg0$): 
Narrow RP pulses ($r$ close to 1) are more efficient than broad pulses to synchronize heterogeneous 
populations of oscillators. Indeed, note that the synchronization boundary of the narrowest pulse 
($r=0.95$) reaches the highest value of the heterogeneity parameter $\Delta$ in Fig.~\ref{fig::q_rp}(b). 
However, a small discrepancy with this previous rule 
was already noticeable in the $q=0$ curve of Fig.~2(a) in~\cite{PM14}. 
Here we revisit that question and find that, as Fig.~\ref{fig::q_rp}(a) shows, 
for negative PRC offsets the discrepancy is even more dramatic: 
the Hopf boundary is far from attaining the largest $\Delta$ value for the narrowest pulse.
Hence, synchronization is not optimal for narrow pulses, but it also depends on the sign of 
the PRC's offset $q$. Consequently, one is tempted to wonder if, 
in nature, adaptation may in some cases drive PRC offsets and pulse widths to be mutually optimized 
in a certain sense.

%%%%%%%%%%%%%%%%%%%%%%%%%%%%%%%%%%%%%%%%%%%%%%%%%%%%%%%%%%%%%%%%%%%%%%%%%%%
%%%%%%%%%%%%%%%%%%%%%%%%%%%%%%%%%%%%%%%%%%%%%%%%%%%%%%%%%%%%%%%%%%%%%%%%%%%

%%%%%%%%%%%%%%%%%%%%%%%%%%%%%%%%%%%%%%%%%%%%%%%%%%%
\subsection{Ariaratnam-Strogatz (AS) pulse} \label{sec::as}
%%%%%%%%%%%%%%%%%%%%%%%%%%%%%%%%%%%%%%%%%%%%%%%%%%%

%%%%%%%%%%%%%%%%%%%%%%%%%%%%%%%%%%%%%%%%%%%%%%%%%%%%%%%%%%%%%%%%%%%%
\begin{figure}
  \includegraphics[width=\columnwidth]{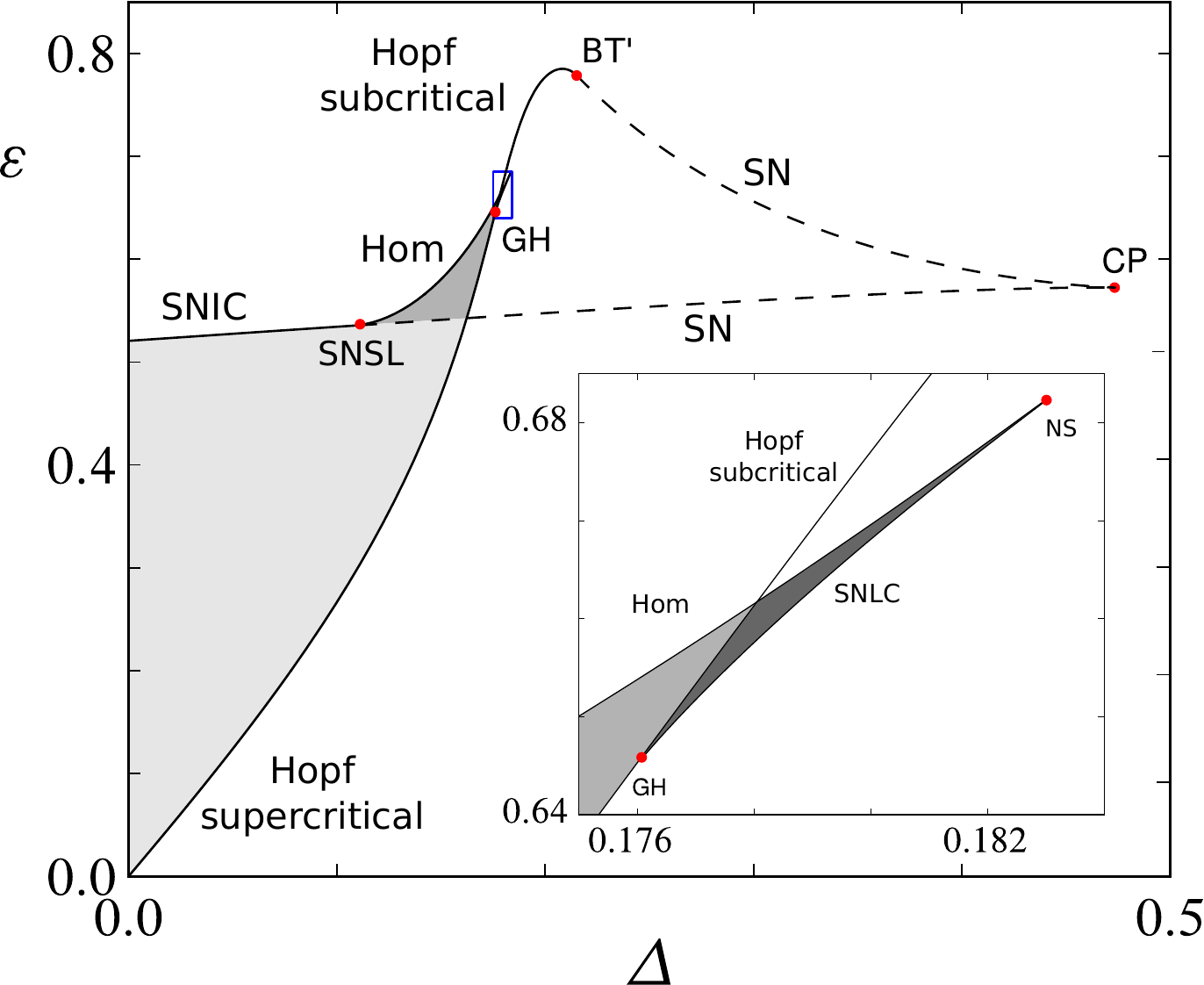}
  \caption{Phase diagram of the Winfree model in the
  $(\Delta,\varepsilon)$-plane for the AS pulse. Parameter values are $q=-1$ and
  $n=5$; the inset is a zoom of the region enclosed by a rectangle in
  the main plot.}
  \label{fig::pdas}
\end{figure}
%%%%%%%%%%%%%%%%%%%%%%%%%%%%%%%%%%%%%%%%%%%%%%%%%%%%%%%%%%%%%%%%%%%%

Thus far we found no qualitative difference between the phase diagram for RP pulses of
Fig.~\ref{fig::pdrp}, with that of Ref.~\cite{PM14}, obtained using AS pulses.  
Nonetheless, in this section we show that this qualitative agreement breaks down for 
AS pulses with PRCs with negative offset.

The AS pulse with $n=1$ is identical to the RP pulse with $r=0$, 
so that no differences arise in this case. 
Surprisingly, when we considered narrower pulses (larger values of 
$n$) a more complicated bifurcation 
scenario showed up, see Fig.~\ref{fig::pdas} for $n=5$ and $q=-1$. 
Indeed, 
at a certain critical $n$, the Bogdanov-Takens point mutates its
character in such a way that the Hopf line 
emanating from it 
becomes of subcritical type, while the homoclinic bifurcation 
now involves an unstable periodic orbit 
---because the sum of the eigenvalues of the saddle point, called saddle quantity,
is positive. This mutated Bogdanov-Takens point is designated as BT' hereafter.
Points BT and BT' are both equally generic zero-eigenvalue points consistent
with the normal form in textbooks \cite{Guckenheimer,Kuznetsov}: BT is the usual
representation (up to a transformation of parameters), while BT' is also
consistent upon time inversion.

In the transition from BT to BT' two new codimension-two 
points appear: 
\begin{enumerate}
\item A generalized Hopf (GH) point on top of the Hopf line where the
bifurcation shifts from super- to sub-critical.
\item  A neutral saddle (NS) point where the homoclinic connection is
degenerate, since it involves a saddle point with zero saddle quantity \cite{Kuznetsov}. At the
NS point the line of homoclinic bifurcation of the {\em stable} limit
cycle terminates.
\end{enumerate}
The GH and NS points are connected by a new line curve, which is the 
locus of a saddle-node bifurcation of limit cycles (SNLC). Figure~\ref{fig::pdsketch}(b) shows
sketches of the phase portraits when a BT' point is present in the phase diagram.
Notably, the synchronization region is detached from the BT' point,
and a region with three attractors (i.e.,~{\em
tristability}) exists.  This region is the approximate triangle with vertices at
GH and NS visible both in the inset of Fig.~\ref{fig::pdas} and in 
Fig.~\ref{fig::pdsketch}(b), region 5.  
There, the limit cycle (corresponding to synchronization) coexist with two stable fixed points.  
Note that by entering into region 5 through the saddle-node bifurcation of limit cycles line 
(connecting the points GH and NS)
a finite-sized limit cycle with a finite basin of attraction suddenly appears.

\subsection{Transition between the synchronization scenarios BT and BT'} \label{sec::bt}
%%%%%%%%%%%%%%%%%%%%%%%%%%%%%%%%%%%%%%%%%%%%%%%%%%%%%%%%

\begin{figure}
  \includegraphics[width=\columnwidth]{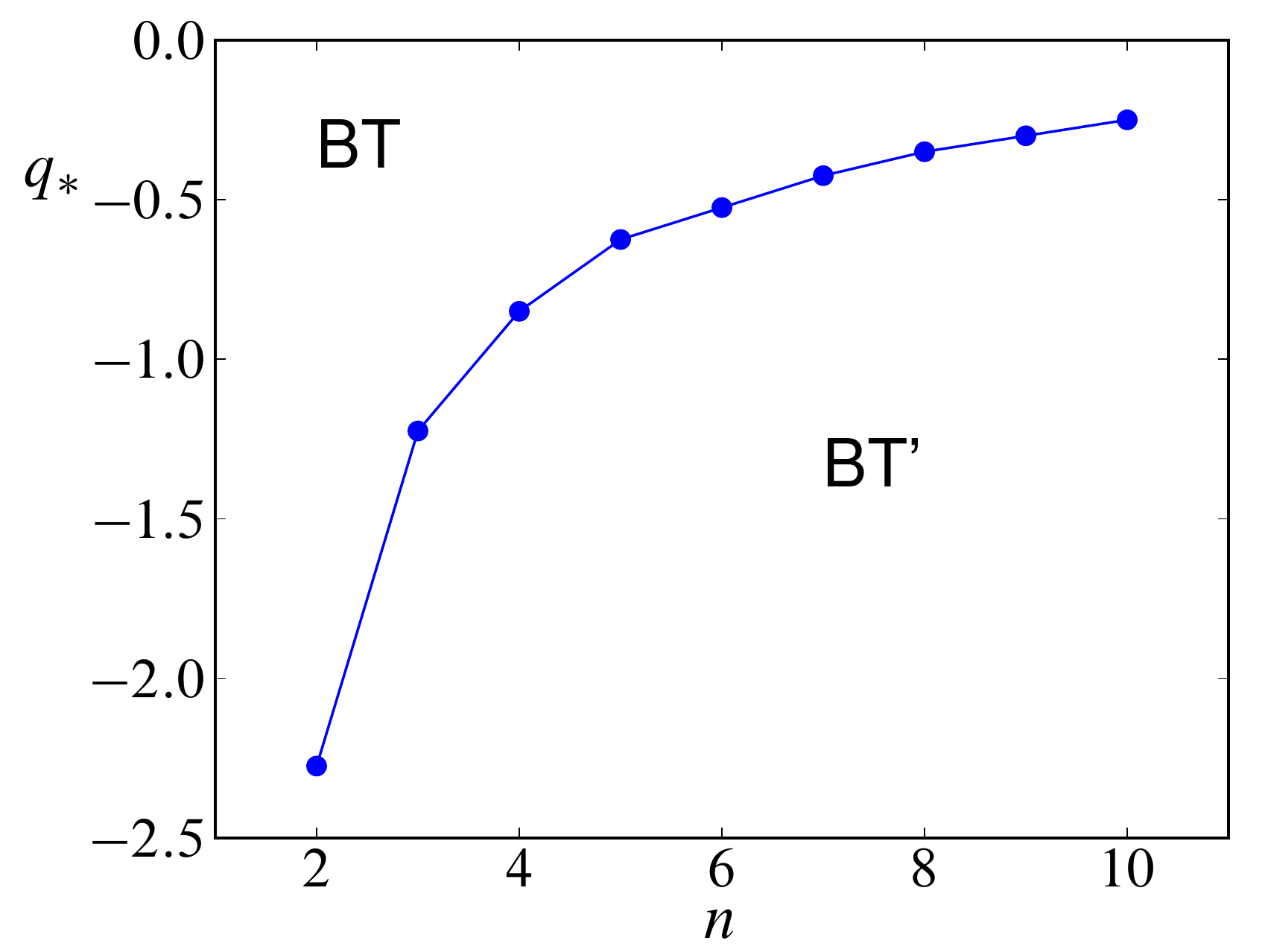}
  \caption{
%%Threshold value of $q^*$ as a function of $n$ 
Numerically obtained critical boundary $q^*$ separating the regions BT and BT',
as a function of the AS pulse width $n$. The novel synchronization scenario,
associated to a BT' point,
shows up for narrow pulses and negative PRC offsets.}
  \label{fig::qn}
\end{figure}

In view of the distinct phase diagrams associated to BT and BT', next 
we investigate the conditions under which each scenario shows up. Our systematic
numerical investigation indicates that the RP pulse is always associated to a BT point. 
In the case of the AS pulse, we determined numerically the
threshold value of $q$, which we designated as $q_*$, where the transition between 
BT and BT' occurs, i.e.~the $q$ value at which a degenerate (codimension-three) 
BT point arises. 
The result covering all integer values $n\le10$
is depicted in Fig.~\ref{fig::qn}, and demonstrates that the BT' point only
arises for sufficiently negative offsets $q$. Noteworthy, when $n$ grows, BT' can be
observed for increasing small values of $|q|$. 
The absence of a point for $n=1$ in Fig.~\ref{fig::qn} is not an omission;
in fact, we failed to numerically find a BT' point even after
considering extremely small values of $q$  
---recall also that the AS pulse with $n=1$ coincides with the RP pulse with $r=0$.
Finally, we carried out 
numerical simulations using square pulses (not shown),
and found that the
bifurcation scenario associated to BT' is observed already for $q=0$, provided
that $b$ is smaller than $b_*(q=0)=1.02\ldots$ (a significantly large value).

%%%%%%%%%%%%%%%%%%%%%%%%%%%%%%%%%%%%%%%%%%%%%%%%%%%%%%%%%%%%%%%%%%%%
%%%%%%%%%%%%%%%%%%%%%%%%%%%%%%%%%%%%%%%%%%%%%%%%%%%%%%%%%%%%%%%%%%%%

To get some more physical insight, 
we examined the asymptotic behavior of $P(\theta)$ in a
neighborhood of $\theta=\pi$ for each pulse type. They are:
$$
P(\pi\pm\delta\theta) \simeq \frac{1-r}{2(1+r)^2}\delta
\theta^2, \frac{(n!)^2}{(2n)!}\delta \theta^{2n}, 0,
$$ 
for RP, AS, and
square pulses, respectively. The marked differences of the respective
asymptotics led us to conjecture a simple rule of the thumb: pulses that fall
fast enough to zero at $\theta=\pi$ are prone to exhibit the synchronization
scenario with five codimension-two points, i.e.~BT'. On the contrary, for pulses
that fall to zero more slowly (such as the RP pulse) favor the first scenario (BT), 
making the second scenario (BT') impossible or only present
for small enough PRC offsets $q$. 

%%%%%%%%%%%%%%%%%%%%%%%%%%%%%%%%%%%%%%%%%%%%%%
\subsection{Dirac delta pulse}\label{sec::dirac}
%%%%%%%%%%%%%%%%%%%%%%%%%%%%%%%%%%%%%%%%%%%%%%

All the 
pulses studied in this paper have the Dirac delta as limiting case. It is not difficult to
obtain the expression of $h(Z)$ for the Dirac delta pulse, see Table I. Nonetheless 
some caution must be taken here: 
for obtaining the mean field $h(Z)$ the thermodynamic limit ($N\to\infty$)
is assumed prior to the zero width pulse limit,
and  it is well known that 
these two limits do not commute \cite{ZLP+07}. 
Therefore, the results we obtain for Dirac delta pulses 
cannot be exactly reproduced in numerical simulations,
which necessarily involve a finite number of oscillators. 
Accordingly, 
the results obtained here for Dirac delta pulses
must be 
interpreted as a limit of the bifurcation lines
for very narrow pulses.
This allows us to put aside the pulse shape, and to focus solely on the
influence of the PRC offset parameter $q$.

\begin{figure}
  \includegraphics[width=\columnwidth]{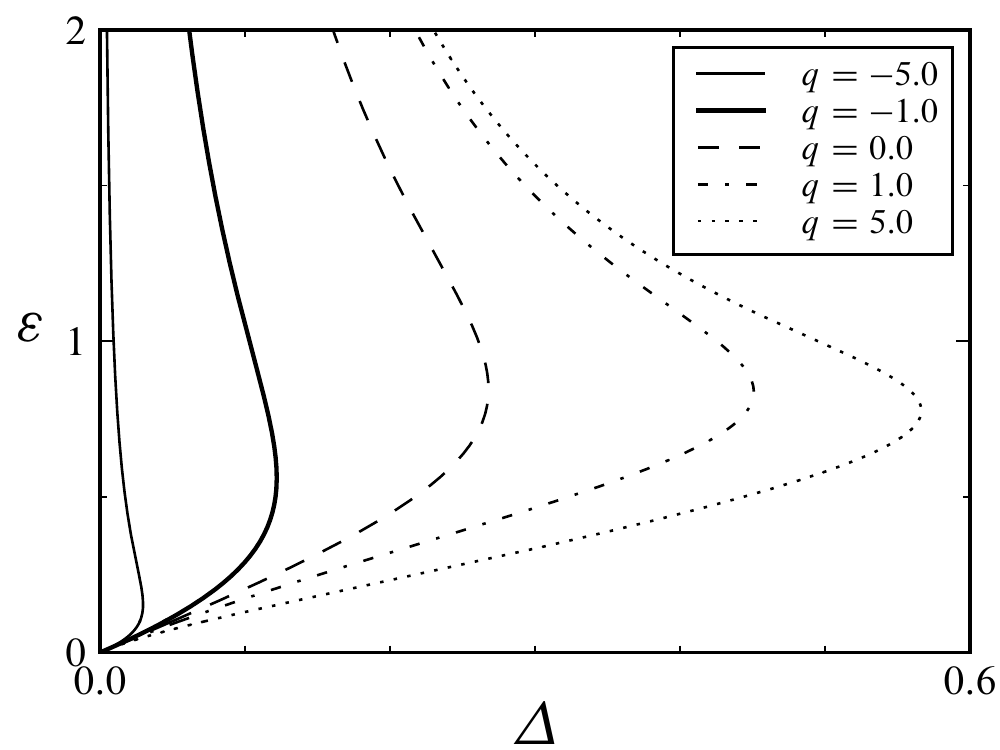}
  \caption{Phase diagram of the Winfree model in the
  $(\Delta,\varepsilon)$-plane for the Dirac delta pulse and several values of
  $q$.}
  \label{fig::deltadirac}
\end{figure}

Figure \ref{fig::deltadirac} shows phase diagrams in the
$(\Delta,\varepsilon)$-plane for Dirac delta pulses, and for several values
of $q$.  The curves displayed are Hopf bifurcation lines that emanate from
the origin and approach the vertical axis when $\varepsilon\to\infty$.
As mentioned above, there are subtle questions regarding this pulse, so the Hopf bifurcation lines
have to be understood simply as the limit of the Hopf curves for very narrow pulses. 
In fact, the absence of the saddle-node bifurcations lines indicates a certainly
singular behavior in that limit.

Yet, from Fig.~\ref{fig::deltadirac}, we can conclude that 
the synchronization region increases monotonically with $q$.
Our physical interpretation of this 
feature is that, for negative PRC offsets,
the bias of the PRC tends to slow down the oscillators favoring the
formation of a cluster with quiescent oscillators (partial oscillation death).
On the contrary, positive PRC offsets generally favor phase advances,
retarding the accumulation of quiescent oscillators, and
leaving room for the synchronization to occur more widely.

Let us finally point out that the bifurcation lines can be analytically obtained
by transforming Eq.~\eqref{Z} into a complex-valued ODE for a new variable $w=(1+Z)/(1+Z)$,
such that $h$ equals $\text{Re}(w)$. In the new coordinate system, and
with the assistance of {\sc mathematica}, we derived convoluted,
but nonetheless exact equations of the Hopf boundaries in parametric form:
\begin{eqnarray}
\Delta_H(y)&=&\frac{f(y,q) \left[-g(y,q)+y (q+y)+1\right] }{\left(y^2+1\right) (2 q+y)}\nonumber \\
\varepsilon_H(y)&=&\frac{2 f(y,q) \left\{g(y,q)(y^2-1)
+y \left[q \left(y^2+3\right)+y\right]+1\right\}}{\left(y^2+1\right) (2 q+y) \left(4q y-y^2+3\right)} \nonumber
\end{eqnarray}
with
\begin{eqnarray}
g(y,q)&\equiv&\sqrt{\left(q^2+1\right) y^2+1} \nonumber\\
f(y,q)&\equiv&\sqrt{2 g(y,q)+2 q y-y^2-1} \nonumber
\end{eqnarray}
where $y\in(0,\infty)$.
In passing, we note, that for $q=0$ a simple explicit formula can be found, see Ref.~[36] in \cite{PM14}.

%%%%%%%%%%%%%%%%%%%%%%%%%%%%%%%%%%%%%%%%%%%%%%%%%%%%%%%%%%%%%%%%%%%%%%%%%%%%%%%%%%%%%%%

%==========================
\section{Limit of weak coupling and nearly identical oscillators}
\label{sec::kuramoto}
%==========================

To conclude, we investigate the Winfree model in the limit of weak 
coupling and weak heterogeneity, i.e.~$|\varepsilon|\ll1$, $\Delta\ll1$.
This is an important limiting case, since 
the method of averaging can be applied 
and the Winfree model reduces to the well-studied Kuramoto-Sakaguchi (KS)
model~\cite{Kur84,SK86}. Our aim in this section is to 
investigate how the synchronization threshold 
of the Winfree model deviates from that of its corresponding KS model, 
for different pulse functions and PRC types.

%%%%%%%%%%%%%%%%%%%%%%%%%%%%%%%%%%%%%%%%%%%%%
\subsection{Averaging approximation: Kuramoto-Sakaguchi model}
\label{sec_av}
%%%%%%%%%%%%%%%%%%%%%%%%%%%%%%%%%%%%%%%%%%%%%

In the classical analysis by Kuramoto \cite{Kur84}, weakly interacting
oscillators with frequencies close to a resonance are described by means of
the averaging approximation. In the case of 1:1 resonance (nearly identical frequencies), 
the interaction 
term between any two oscillators becomes a function of their phase difference.

Given that the PRCs 
considered here 
are chosen to be sine-shaped, the only resonant term is
the first harmonic. Thus, the 
averaging 
calculation 
leads to the KS model \cite{Kur84,SK86,PM14}:
\begin{equation}
\dot \theta_i=\omega_i +\varepsilon 
q+ \Pi  \frac{\varepsilon}{N}
\sum_{j=1}^N\left [\sin(\theta_j-\theta_i)
-q \cos(\theta_j-\theta_i) \right]  
\label{av_winfree}
\end{equation}
The parameter $\Pi$ is a ``shape factor'' that depends on the pulse shape.
The shape factor depends only  on the first harmonic of the pulse, 
or more precisely, $\Pi=c_1$, see Eq.~\eqref{prc_fourier}. 
The dependence of $\Pi$ on the parameter controlling the
pulse width can be found in the last column of Table \ref{table}.
In all cases, the effective interaction strength increases 
as the pulses become narrower. In fact, the largest $\Pi$ value is attained for the Dirac delta pulse.

Generally, for
unimodal frequency distributions, the KS model Eq.~\eqref{av_winfree} displays a 
simple transition between incoherence (asynchrony)
and macroscopic synchronization at a critical finite value of $\varepsilon$
---but see \cite{OW12,OW16} for exceptions.
The synchronous state is characterized by the appearance of a subset of oscillators
that rotate with a common frequency and have their phases locked, thanks to
the mutual coupling that is able to overcome the disparity of the natural frequencies.
For the Lorentzian distribution of frequencies, Eq.~\eqref{lorentzian}, 
the critical coupling of the synchronization transition 
in the thermodynamic limit ($N\to\infty$)
can be obtained analytically \cite{SK86}:
\begin{equation}
\varepsilon_c^{(\text{av})}= \frac{2\Delta}{\Pi},
\label{av_c}
\end{equation}
where the superscript ``($\text{av}$)'' is used to emphasize
that this is the critical coupling of the averaged model in Eq.~\eqref{av_winfree}.
Curiously, within this approximation $\varepsilon_c$ does not depend on $q$. (This 
has to be attributed to the special properties of the Lorentzian distribution $g(\omega)$
which uses to yield particularly simple results in Kuramoto-like models.) 
The exact critical coupling of the Winfree model is
computed numerically below and, as presumed, depends on $q$.

%%%%%%%%%%%%%%%%%%%%%%%%%%%%%%%%%%%%%%%%%%%%%%%%%%%%%%%%%%%%%%%%%%%%%%%%%%%%
\subsection{Synchronization threshold: Winfree vs. KS model}
%%%%%%%%%%%%%%%%%%%%%%%%%%%%%%%%%%%%%%%%%%%%%%%%%%%%%%%%%%%%%%%%%%%%%%%%%%%%%

%%%%%%%%%%%%%%%%%%%%%%%%%%%%%%%%%%%%%%%%%%%%%%%%%%%%%%%%%%%%%%%%%%%%%%%%%%%%%%
\begin{figure}%[t]
  \includegraphics[width=\columnwidth]{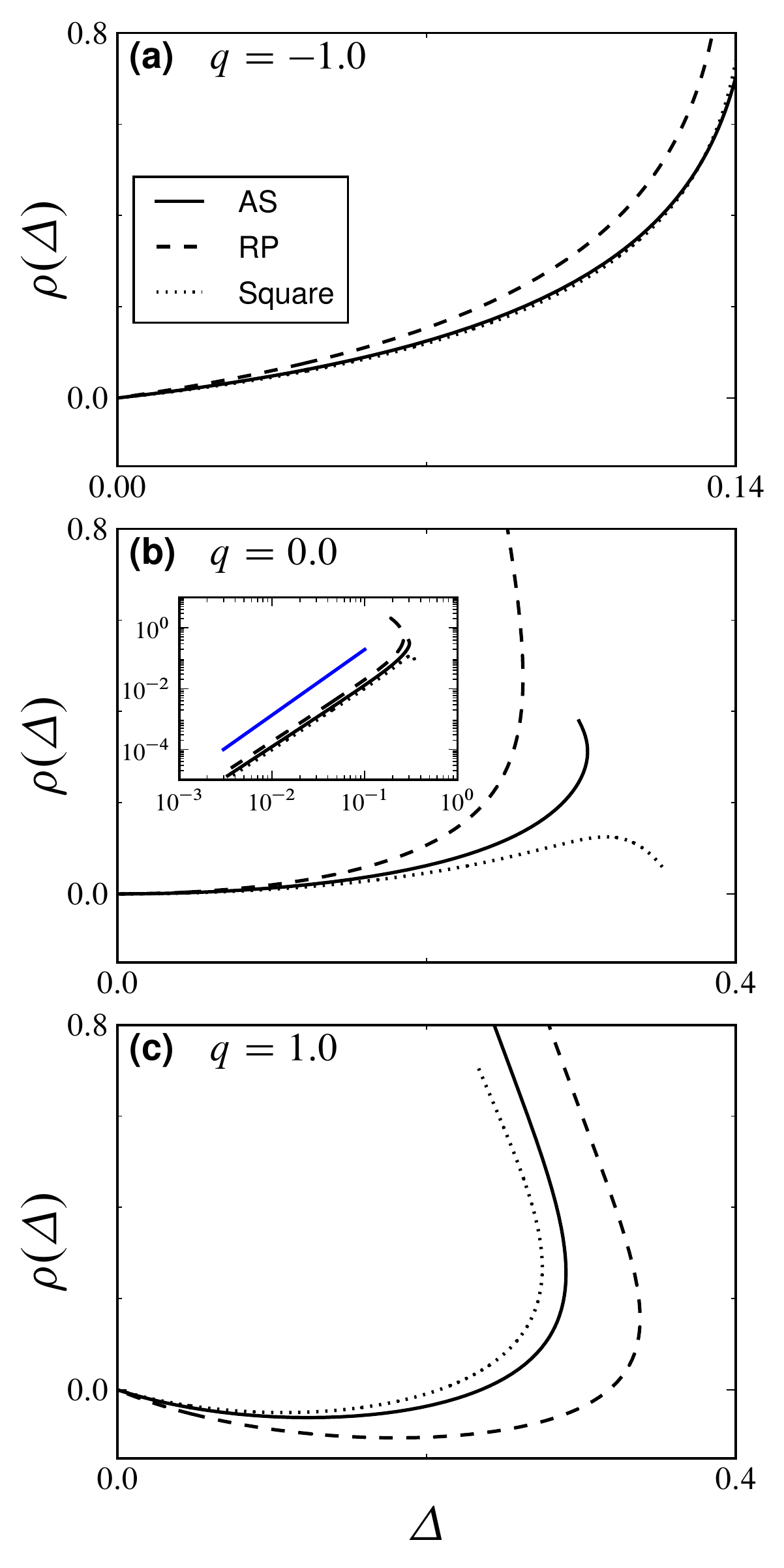}
  \caption{
Quantity $\rho(\Delta)$, Eq.~\eqref{eq::rho}, measuring the deviation from the
averaging approximation Eq.~\eqref{av_winfree} of the Winfree model for AS,
RP, and square pulses with $\Pi=10/11$ ---see Table \ref{table}. Panels (a),
(b) and (c) correspond to $q=-1$, $q=0$ and $q=1$, respectively. The inset in
panel (b) shows the curves in log-log scale. The thick solid line has slope 2
and is plotted as a guide for the eye.}
  \label{fig::KSmodel}
\end{figure}
%%%%%%%%%%%%%%%%%%%%%%%%%%%%%%%%%%%%%%%%%%%%%%%%%%%%%%%%%%%%%%%%%%%%%%%%%%%%%%

To test the goodness of the averaging approximation we compare the synchronization threshold
of the Winfree model with the threshold of
its averaged counterpart, given by Eq.~\eqref{av_c}.  Our aim is to
determine if certain pulses deviate more from the
averaging approximation, and whether these results depend on the PRC offset $q$.
To make the comparison significant we considered different pulse types with 
the same $\Pi$ values. 
In different panels of Fig.~\ref{fig::1}, pulses plotted with the same line style
have identical shape
factor $\Pi$, and therefore they yield the identical KS model upon averaging.
In turn, the prediction of Eq.~\eqref{av_c} is exactly the same for all pulse types 
(provided the same $\Pi$ value). 
In order to measure the deviation of the Winfree model from the 
the KS model we define the quantity
\begin{equation}\label{eq::rho}
\rho(\Delta)=\frac{\varepsilon_H-\varepsilon_c^{(\text{av})}}{\varepsilon_c^{(\text{av})}} 
\end{equation}
which is proportional to the difference between the exact and the approximated critical couplings
(normalized by the approximated critical coupling).
For each pulse type and $q$ value, the locus of the Hopf bifurcation $\varepsilon_H(\Delta)$
is numerically available from the exact low-dimensional Eq.~\eqref{Z}.

In Fig.~\ref{fig::KSmodel} we graph $\rho$
for $\Pi=10/11$ and the three pulse types considered, adopting three values $q=-1$, $0$, and $1$
in panels (a), (b) and (c), respectively.
As expected $\rho(\Delta\to0)=0$, indicating 
the validity of the averaging approximation in this limit.
As $\Delta$ is increased from zero
$\rho(\Delta)$ becomes positive for $q=-1$ and negative for $q=1$,
which implies 
that synchronization is hindered (promoted)
with respect to the averaging approximation for negative (positive) $q$.
(This is also consistent with Fig.~\ref{fig::deltadirac}.)
Numerical evidence shows that
\begin{gather*}
  \rho(\Delta)=\xi(q)\Delta+O(\Delta^2)
\end{gather*}
where $\xi(q)$ is a pulse-dependent odd function (and possibly monotonically decreasing).
Note that this means that the best pulse type, in a certain sense, for $q>0$,
becomes the worst for $q<0$. For instance, for $q=-1$ 
synchronizability is the best for the square pulse among the pulses considered, 
but this becomes just the opposite for $q=1$.
The case $q=0$ (zero PRC offset) is the 
boundary between 
these scenarios,
because $\xi(0)=0$. Accordingly, 
the inset of Fig.~\ref{fig::KSmodel}(b) confirms a nonlinear dependence,
namely quadratic, of $\rho(\Delta\ll1)$.

%==========================
\section{Conclusions}\label{sec::con}
%==========================

The Winfree model is broadly known, but scarcely studied in detail. 
The application of the OA ansatz 
to the Winfree model allows for the detailed investigation of its collective dynamics. 
Here we systematically investigated the dynamics of the 
Winfree model for three different pulse types and various widths, 
and for sinusoidal PRCs with positive, 
negative, and zero offsets $q$.

The case of negative offset was not considered in \cite{PM14}, but has revealed to be interesting and
nontrivial. The claim that narrow pulses 
are optimal for synchronizing large populations of oscillators~\cite{PM14}, does not hold 
for negative PRC offsets. In this case we observe that the optimal pulse, allowing  to 
synchronize ensembles with a higher degree of heterogeneity, 
has an intermediate width, see Fig.~\ref{fig::q_rp}(a).
Moreover, for negative offsets (but not only)
it is more likely to find a synchronization scenario with five codimension-two points (incl.~BT'),
in contrast to the scenario with three points reported in \cite{PM14}. Under which
conditions each scenario is found depends on the particular pulse type.
From our results we inferred that pulses which are closer to zero at phases far from the
peak phase are more likely to exhibit a BT' point. In fact, the RP pulse does not exhibit a
BT' point for any $q$ value, while the square pulse already does for a balanced PRC ($q=0$).
We also considered the limit of infinitely narrow pulses (Dirac delta pulses) 
and provided exact formulas for the
synchronization boundary (a Hopf bifurcation). Additionally, we demonstrated 
that positive PRC offsets display 
larger synchronization regions and are capable of synchronizing more heterogeneous ensembles.

Finally, 
we have compared the synchronization threshold of the
Winfree model with its averaging approximation (the Kuramoto-Sakaguchi model), and 
found what we may summarize as antisymmetry with the PRC offset parameter $q$. 

In future studies, it would be interesting to find techniques 
to efficiently analyze the Winfree model with  
frequency distributions beyond the Lorentzian one. 
The study in~\cite{AS01} for a uniform distribution of natural frequencies is valuable, 
but it is difficult to extend it to nonvanishing $q$.
Generalizing the model by considering other sources of heterogeneity is also 
an interesting  venue for future research.

\acknowledgments
We acknowledge support by MINECO (Spain) under Projects 
No.~FIS2014-59462-P, No.~FIS2016-74957-P, No.~PSI2016-75688-P and No.~PCIN-2015-127. 
E.M and D.P also acknowledge support 
by the European Union's Horizon 2020 research and innovation
programme under the Marie Sk{\l}odowska-Curie grant agreement No.~642563.

\bibliography{bibliografia}

\end{document}